\documentclass[a4paper]{spie}  

\usepackage{amsmath,amsfonts,amssymb}
\usepackage{graphicx}

\input{aas_macros.sty}

\usepackage{xspace}
\usepackage[caption=false]{subfig}
\DeclareCaptionListOfFormat{myparens}{#2}
\newcommand{\Subref}[1]{\protect\subref{#1}}
\usepackage[dvipsnames,usenames]{color}

\usepackage[breaklinks={true}, bookmarks, colorlinks={true}, allcolors=blue, pdfpagemode={NON}, pdffitwindow=true, pdfstartview={FitB}, pdftitle={Prototyping the graphical user interface for the operator of CTA}, pdfauthor={Iftach Sadeh}, pdfsubject={Prototyping the graphical user interface for the operator of the Cherenkov Telescope Array}]{hyperref}

\let\orgautoref\autoref
\providecommand{\Autoref}
        {\def\equationautorefname{Equation}%
         \def\figureautorefname{Figure}%
         \def\subfigureautorefname{Figure}%
         \def\chapterautorefname{Chapter}%
         \def\sectionautorefname{Section}%
         \def\subsectionautorefname{Section}%
         \def\subsubsectionautorefname{Section}%
         \def\Itemautorefname{Item}%
         \def\tableautorefname{Table}%
         \def\appendixautorefname{Appendix}%
         \orgautoref}

\renewcommand{\autoref}
        {\def\equationautorefname{Eq.}%
         \def\figureautorefname{Fig.}%
         \def\subfigureautorefname{Fig.}%
         \def\chapterautorefname{Ch.}%
         \def\sectionautorefname{Sect.}%
         \def\subsectionautorefname{Sect.}%
         \def\subsubsectionautorefname{Sect.}%
         \def\Itemautorefname{item}%
         \def\tableautorefname{Table}%
         \orgautoref}

\providecommand{\autorefs}
        {\def\equationautorefname{Eqs.}%
         \def\figureautorefname{Figs.}%
         \def\subfigureautorefname{Figs.}%
         \def\chapterautorefname{Chs.}%
         \def\sectionautorefname{Sects.}%
         \def\subsectionautorefname{Sects.}%
         \def\subsubsectionautorefname{Sects.}%
         \def\Itemautorefname{items}%
         \def\tableautorefname{Tables}%
         \orgautoref}

\newcommand{\ie}{i.e.\/,\xspace}

\newcommand{\eg}{e.g.\/,\xspace}

\newcommand{\djs}{\texttt{d3.js}\xspace}
\newcommand{\dcs}{\texttt{dc.js}\xspace}
\newcommand{\jvs}{\texttt{Javascript}\xspace}
\newcommand{\crsfilt}{\texttt{Crossfilter}\xspace}
\newcommand{\python}{\texttt{Python}\xspace}
\newcommand{\pyramid}{\texttt{Pyramid}\xspace}
\newcommand{\webSock}{\texttt{Web Sockets}\xspace}
\newcommand{\polymer}{\texttt{Polymer}\xspace}
\newcommand{\matDsgn}{\texttt{Material Design}\xspace}
\newcommand{\webComp}{\texttt{Web Component}\xspace}
\newcommand{\google}{\texttt{Google}\xspace}

\title{Prototyping the graphical user interface for the operator of the Cherenkov Telescope Array}

\author[a]{I.~Sadeh}
\author[a]{I.~Oya}
\author[b]{J.~Schwarz}
\author[c]{E.~Pietriga}
\author[d]{the CTA Consortium}
\affil[a]{~DESY-Zeuthen, D-15738 Zeuthen, Germany}
\affil[b]{~INAF - Osservatorio Astronomico di Brera, Italy}
\affil[c]{~INRIA Saclay - Ile de France, LRI (Univ. Paris-Sud \& CNRS), France}
\affil[d]{~\href{http://www.cta-observatory.org/}{http://www.cta-observatory.org/}}

\begin{document} 
\maketitle

\begin{abstract}
The Cherenkov Telescope Array (CTA) is a planned gamma-ray observatory. CTA will
incorporate about 100 imaging atmospheric Cherenkov telescopes (IACTs) at a Southern
site, and about 20 in the North. Previous IACT experiments have used up to five telescopes.
Subsequently, the design of a graphical user interface (GUI) for the operator of CTA
involves new challenges. We present a GUI prototype, the concept for which is being
developed in collaboration with experts from the field of Human-Computer Interaction.
The prototype is based on Web technology; it incorporates a \python web server,
\webSock and graphics generated with the \djs \jvs library.
\end{abstract}



\section{INTRODUCTION}\label{sec_intro}  
%
The Cherenkov Telescope Array (CTA)~\cite{2011ExA....32..193A,actlProcc,actlArcProcc} is a planned observatory
for very high-energy gamma-rays, sensitive to energies from $20$~GeV to a few hundred~TeV.
Gamma-rays induce particle cascades in the atmosphere. These are accompanied by Cherenkov radiation,
which is emitted by the charged particles in the cascade.
The Cherenkov light may be detected by imaging atmospheric Cherenkov telescopes (IACTs)~\cite{Hillas201319}.
Using multiple telescopes in concert, Cherenkov showers can stereoscopically be sampled,
allowing to reconstruct the properties of the primary gamma-ray.

CTA will include three different telescope types,
sensitive to different gamma-ray energy ranges.
The telescopes will be deployed in two sites, Northern and Southern,
which will respectively include about~$100$ and about~$20$ telescopes.
Currently running IACT experiments
such as H.E.S.S.~\cite{2006A&A...457..899A}, MAGIC~\cite{Albert:2007xh}
and VERITAS~\cite{Holder:2006gi},
are restricted to up to five telescopes.
Compared to these instruments, the large number of CTA telescopes
will improve the sensitivity and the energy coverage of gamma-ray measurements by
at least an order of magnitude.

The large number of telescopes has implications for the development of a
graphical user interface (GUI) for the operator of a CTA site. The complexity
of the system presents new and interesting challenges, requiring innovative design.
In the following, we detail the development process of the operator GUI. We
discuss the requirements we have derived for the GUI, and describe our
prototype implementation.\footnote{~Media resources illustrating features of the prototype are available at \url{https://www-zeuthen.desy.de/~sadeh/}~.}

\section{DEVELOPMENT PROCESS AND REQUIREMENTS FOR THE GUI}\label{sec_req}
%
In order to develop an effective user interface, we have drawn upon the lessons learned
from the Atacama Large Millimeter/submillimeter Array (ALMA)~\cite{2009IEEEP..97.1463W}.
ALMA is an astronomical interferometer of radio telescopes in the Atacama desert of northern Chile.
It will be comprised of up to~$66$ radio antennas,
and so will be comparable to CTA in complexity.%

During the early stages of ALMA, conventional user interfaces were developed.
Early experience operating the ALMA array with only a few antennas indicated
that the latter were not adequate.
The initial interfaces required many unnecessary interactions to access
relevant information. This resulted in extraneous cognitive load,
and was not efficient for quick diagnosis of system problems.
The implementation was therefore improved, taking into account
advances in the field of Human-Computer Interaction (HCI)~\cite{pietriga:hal-00735792}.

For creating the GUI for the operator of CTA, we are following the
design process used for ALMA.
The GUI is being developed by involving experienced telescope
operators and astroparticle physicists on the one hand, and experts from the field of HCI on the other.
In contrast to ALMA, we have been able to adopt this model at
the start of the design process for CTA. We will therefore be able to apply HCI input to all aspects of the
operator interface. The development process includes
practical face-to-face meetings,
in tangent with brainstorming workshops with representatives of the relevant stakeholders.
To date, we have held two such participatory design workshops. The outcome of the meetings is twofold.
For one, the workshops helped us to refine the scope of the operator GUI, \ie answer
the question, ``what should the GUI enable users to do?''.
Additionally, we have defined a preliminary set of panels for the GUI. That is,
we have began to address the question, ``how should the GUI be designed?''.


With regards to the first question, we first note that
on-site operations related to observing with CTA will normally be automated.
Consequently, the GUI for the operator of a CTA site will nominally be used to perform the
following tasks:
\begin{enumerate}{}
\item
   initiate and end observations;
\item
   override the automated scheduled operations in order to perform a specific
   task, or for safety reasons, which might require manual control over a given telescope or group
   of telescopes;
\item
   monitor the state of the array during data acquisition, which includes monitoring
   of low-level hardware components, of software processes, and of the output
   of a near real-time data analysis;
\item
   identify and diagnose problems with specific sub-systems or processes, in order to solve minor problems
   or to notify technical experts, as needed.
\end{enumerate}

With regards to the second question, we have defined the following categories,
which will be represented by different GUI panels (some of which may overlap):
\begin{enumerate}{}
\item
\textbf{Telescope:} monitor the status of telescopes and their respective sub-systems;
manually control single telescopes and sub-arrays, etc.
   
\item
\textbf{Process:} monitor and modify predefined operation sequences, such as
starting up the array at the beginning of the night, and performing a
scheduled observing task.
   
\item
\textbf{Science:} monitor the output of a physics analysis on the level
of a single telescope or of a sub-array, including trigger rates, event-reconstruction metrics,
science summary reports, etc.
   
\item
\textbf{Infrastructure:} monitor the status of auxiliary systems, such
as the power grid, as well as other indicators, such as
alarms and data transfer rates.

\item
\textbf{Environment:} monitor environmental systems, such as
weather monitors, all-sky cameras, etc.

\item
\textbf{Miscellaneous:} provide access to terminals, shift logs, expert call sheets, etc.
\end{enumerate}

With these guidelines in mind, we have identified the following set of initial requirements for the GUI:
\begin{enumerate}{}
\item
   the GUI will be able to convey information for various levels of telescope multiplicity; on the
   level of the entire array ($\sim100$~telescopes); the level of sub arrays (between~$1$ and~$32$ groups); or
   the level of a single telescope and the associated sub-systems.
   Multi-telescope views for a given sub-system will also be available,
   which will take into account differences between telescope types.
   The various levels of information will be integrated using semantic zooming (defined below);
\item
   the GUI will need to be highly responsive, 
   where the different panels of the GUI will have the option to be synchronized;
\item
   the GUI will integrate different interfaces to CTA
   software (databases, live feeds etc.) with various latencies;
\item
   the different components of the GUI will be modular, allowing for quick updates and for collaborative development.
\end{enumerate}

In the next section, we illustrate the requirements for the operator GUI
using a prototype design.
At this stage of development, the prototype by no means addresses the complete scope of the GUI.
Instead, it acts as a proof of concept for those features which were identified as most important.

\section{PROTOTYPING ACTIVITIES}\label{sec_req}
%
Our prototype is based on Web technologies.
A Web-based framework has the advantage of being lightweight and modular. It also naturally
allows for remote access, which will enable the GUI to be used for remote monitoring,
and for remote operation. These will be useful both for physicists and
for technical experts, who the operator of the array will need to consult.

The back-end of the prototype is a \python server called
\pyramid.\footnote{~See \href{http://docs.pylonsproject.org/projects/pyramid/}{http://docs.pylonsproject.org/projects/pyramid/}~.}
A \python framework has the advantage of being very versatile, incorporating the
ability to use many off-the-shelf libraries for computation, communication,
multi-threading and multi-processing.
The front-end is a web browser. The design is responsive, \ie adjustable according to the width
of the display. It is implemented using \polymer,\footnote{~See \href{https://www.polymer-project.org/}{https://www.polymer-project.org/}~.}
a \webComp application programming interface (API),
developed by \google, based on
the \matDsgn\footnote{~See \href{https://design.google.com/}{https://design.google.com/}~.} concept.
Data are displayed using an open-source \jvs library,
called \djs.\footnote{~See \href{https://d3js.org/}{https://d3js.org/}~.}
The latter is a data-driven framework, with integrated mechanisms for displaying,
updating and animating vector graphics.  Communication between the back-end and the
front-end of the GUI is performed using \webSock. These allow asynchronous communication,
facilitating quick and robust GUI behaviour. 
For plotting, we also use 
\dcs,\footnote{~See \href{https://github.com/dc-js/dc.js}{https://github.com/dc-js/dc.js}~.}
a library based on \djs and on
\crsfilt.\footnote{~See \href{http://square.github.io/crossfilter/}{http://square.github.io/crossfilter/}~.}
The latter is a \jvs library, which facilitates exploring large multivariate datasets.

\begin{figure*}[htbp]
\begin{center}
  \begin{minipage}[c]{0.5\textwidth}
    \centering\subfloat[]{\label{FIG_arrZoom_0}\includegraphics[trim=0mm -30mm 0mm -25mm,clip,width=.95\textwidth]{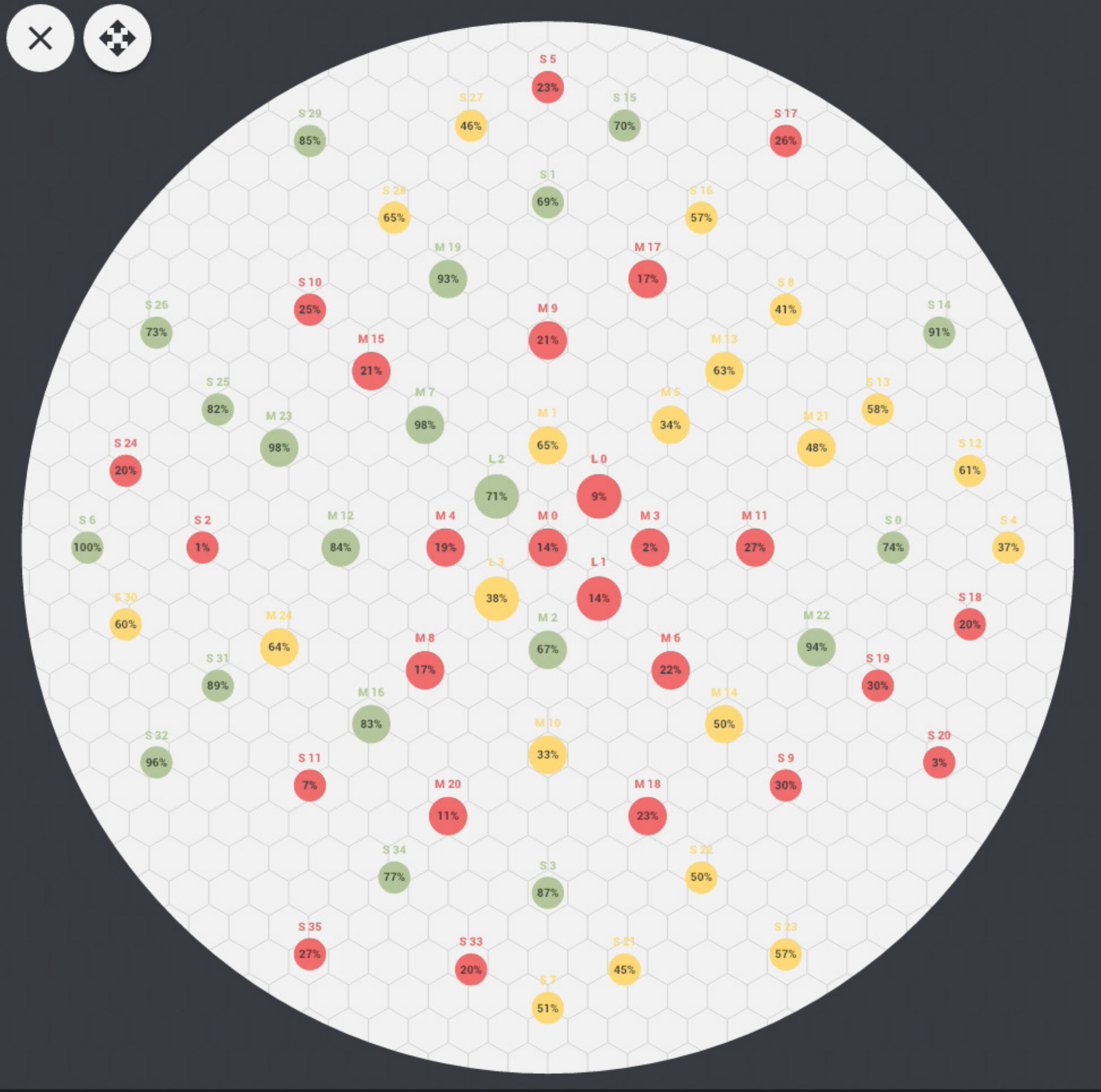}}
  \end{minipage}\hfill
  \begin{minipage}[c]{0.5\textwidth}
    \centering\subfloat[]{\label{FIG_arrZoom_1}\includegraphics[trim=0mm -30mm 0mm -25mm,clip,width=.95\textwidth]{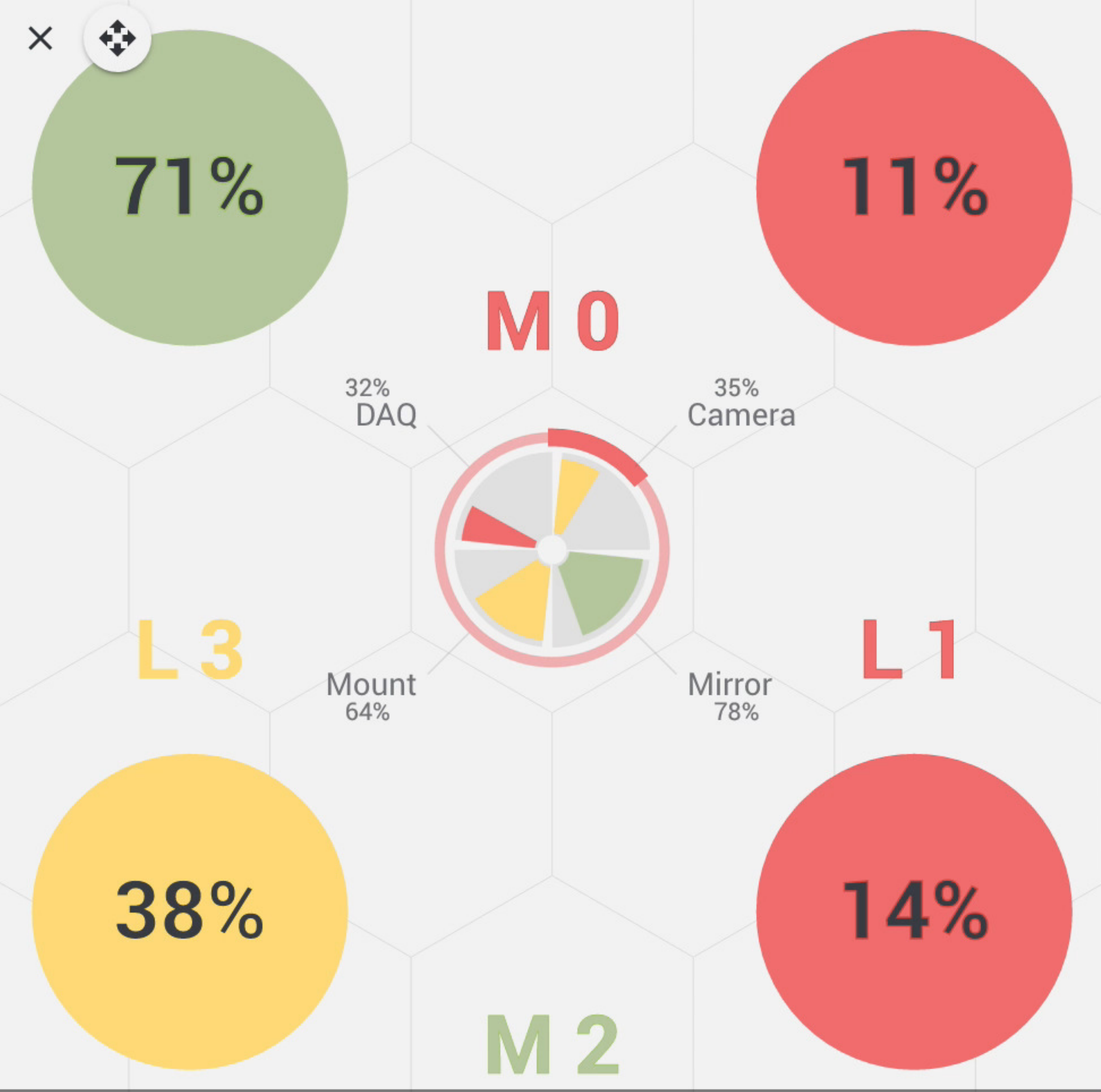}}
  \end{minipage}\hfill
  \begin{minipage}[c]{0.5\textwidth}
    \centering\subfloat[]{\label{FIG_arrZoom_2}\includegraphics[trim=0mm -30mm 0mm -25mm,clip,width=.95\textwidth]{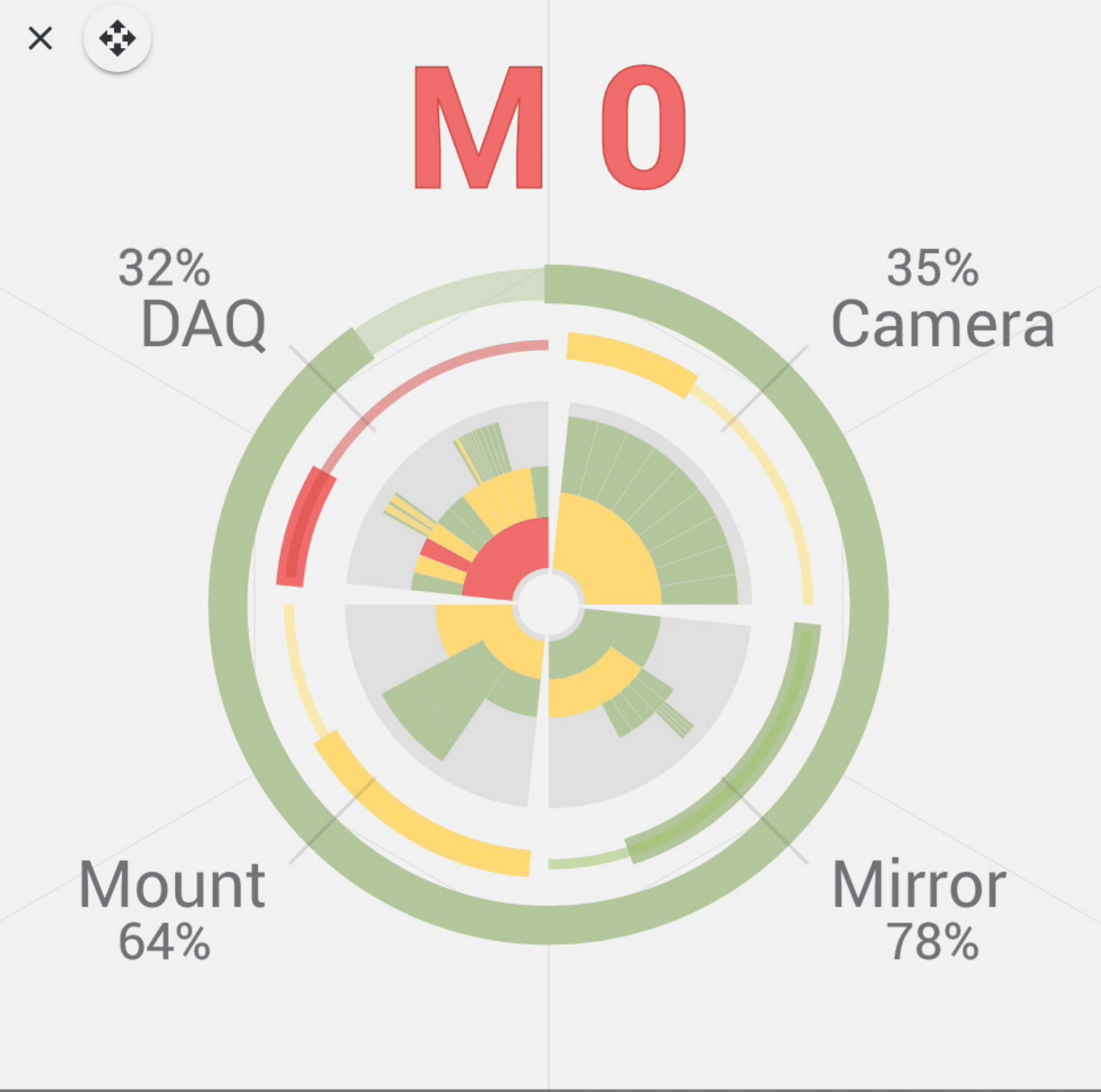}}
  \end{minipage}\hfill
  \begin{minipage}[c]{0.5\textwidth}
    \centering\subfloat[]{\label{FIG_arrZoom_3}\includegraphics[trim=0mm -30mm 0mm -25mm,clip,width=.9425\textwidth]{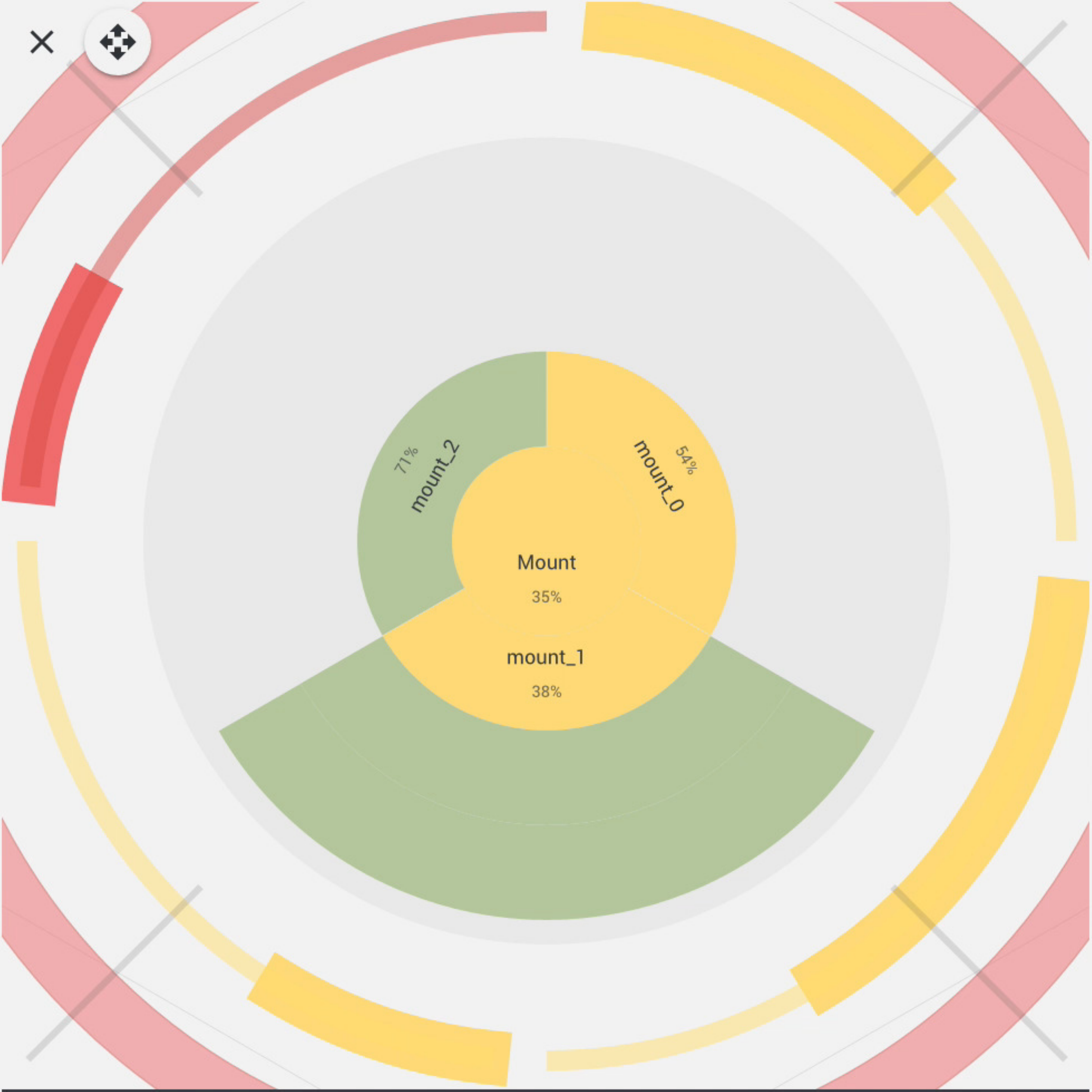}}
  \end{minipage}\hfill
  \vspace{15pt}
  \caption{\label{FIG_arrZoom}
     Pseudo-geographic display of a CTA array, showing different semantic zoom levels, as described in the text.
     The zoom factors are~$1$ in~\Subref{FIG_arrZoom_0}, $7$ in~\Subref{FIG_arrZoom_1},
     and~$\sim14$ in~\Subref{FIG_arrZoom_2} and~\Subref{FIG_arrZoom_3}.
  }
\end{center}
\end{figure*}

\Autoref{FIG_arrZoom} shows a representation of the layout of a CTA array, where each circle corresponds to
a single telescope.
The relative position of telescopes corresponds to a scaled physical layout
on-site, constituting a pseudo-geographic display of the array.
The attached numerical values and the
colour scheme of elements correspond to a \textit{health metric}. This metric
may represent different properties according to the \textit{visualization context}.
The panel incorporates \textit{semantic zooming}~\cite{Perlin1993}, which defines the context.
Semantic zooming is a technique
of assigning different layers of information to a given visual element. This
behaviour complements the usual geometric zooming,
where the size of elements changes with the zoom factor.
For the type of semantic zoom implemented in this example,
the level of detail associated with an element increases as one zooms-in.
The figure shows the evolution of the display for different zoom factors, as follows.
%

\textbf{Zoom factor~$1$:}
   \Autoref{FIG_arrZoom_0} shows a global view of all telescopes, emphasizing
   physical positions, as well as a general health metric. In this example,
   health corresponds to the combined health of all sub-systems of the telescope.
   For illustration, one of these sub-system, the camera, reports reduced health as the
   percentage of dead pixels.

\textbf{Zoom factor~$7$:}
   Given this zoom level, \autoref{FIG_arrZoom_1} shows how
   the display changes for a given element which is hovered over.
   The hover action triggers a transition, where a circle representing a single telescope
   is replaced by a more detailed view of the associated sub-systems.
   The context from the previous zoom level is preserved during the transition. The
   health metric, which was previously represented by the full circle, has morphed
   into the red circle surrounding the four wedges.
   Each of the wedges represents one of the sub-systems of a telescope:
   the data-acquisition system (DAQ);
   the camera; the mirror system; and the mount (the structure of the telescope). Each of
   the latter now also reports its individual health metric. These are
   represented numerically, by colour and by the percentage of filled area
   within the corresponding wedge.

\textbf{Zoom factor~$\sim14$:}
   Given this zoom level, another transition occurs. \Autoref{FIG_arrZoom_2} shows
   a second level of components, which is exposed for each of the
   four sub-systems. As before, the context from the previous zoom level is preserved;
   the health metric of each of the sub-systems is now represented by a thin outer arc.
   The inner wedges now each have different layouts of components. These are arranged in
   \textit{sunburst diagrams}~\cite{Stasko00}, which are a hierarchical variation on a pie chart.
   The change from the view shown in \autoref{FIG_arrZoom_2} to that shown in \autoref{FIG_arrZoom_3}
   occurs as the user clicks on the wedge of the mount. The sunburst diagram
   of the latter then opens up to a full~$360^{\circ}$, and the wedges of the DAQ, camera and mirror
   disappear. The generic names (\texttt{mount\_0}, \texttt{mount\_1}, etc.)
   are temporary place-holders for different elements of the mount sub-system,
   such as drivers and PLC (programmable logic controller)~\cite{2012SPIE.8451E..0HB} status indicators.
%

\begin{figure*}[htbp]
\begin{center}
  \begin{minipage}[c]{1\textwidth}
    \centering\subfloat[]{\label{FIG_arrPos_0}\includegraphics[trim=0mm -30mm 0mm -25mm,clip,width=.68\textwidth]{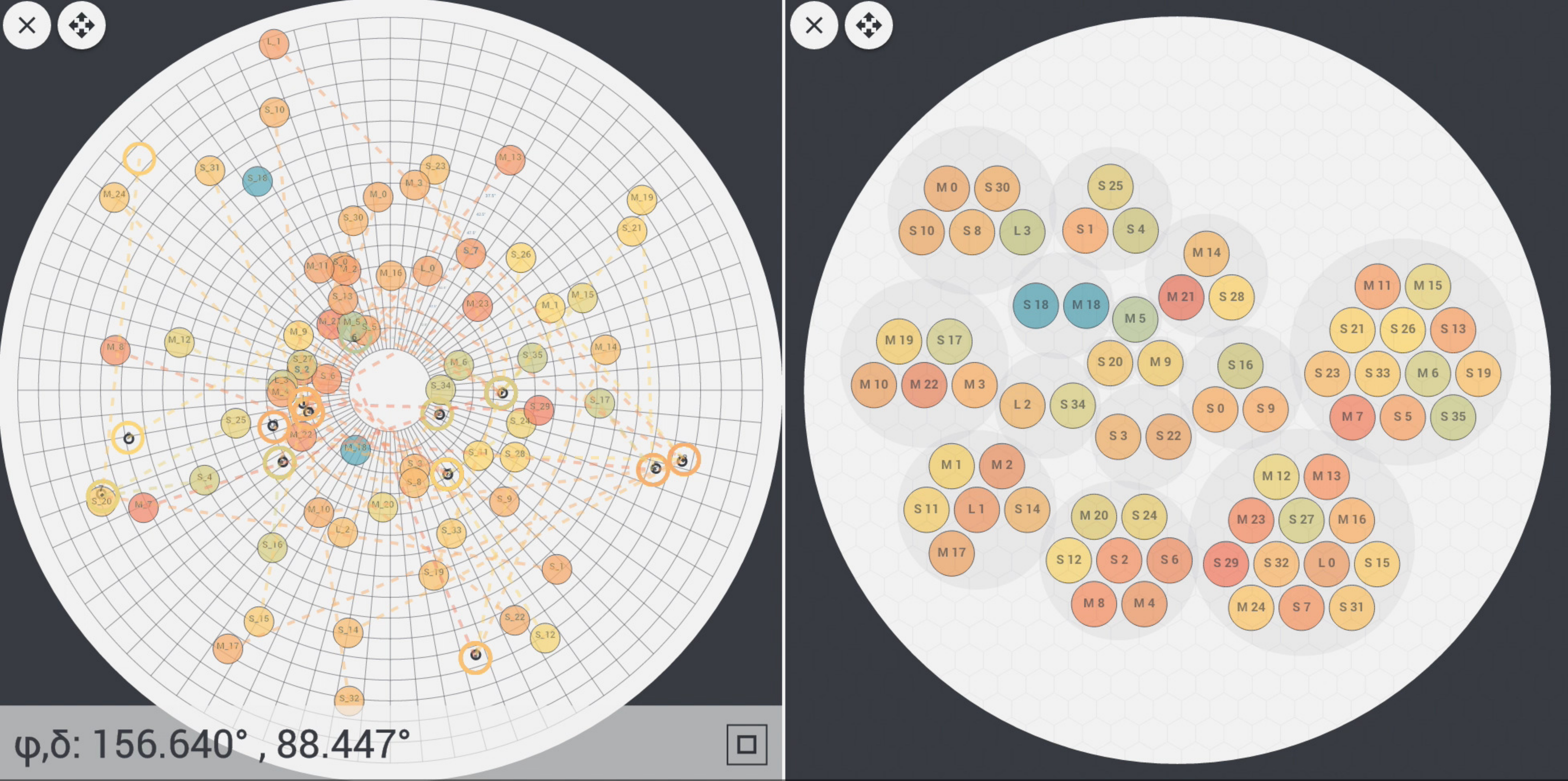}}
  \end{minipage}\hfill \\
  \begin{minipage}[c]{1\textwidth}
    \centering\subfloat[]{\label{FIG_arrPos_1}\includegraphics[trim=0mm -30mm 0mm -25mm,clip,width=.68\textwidth]{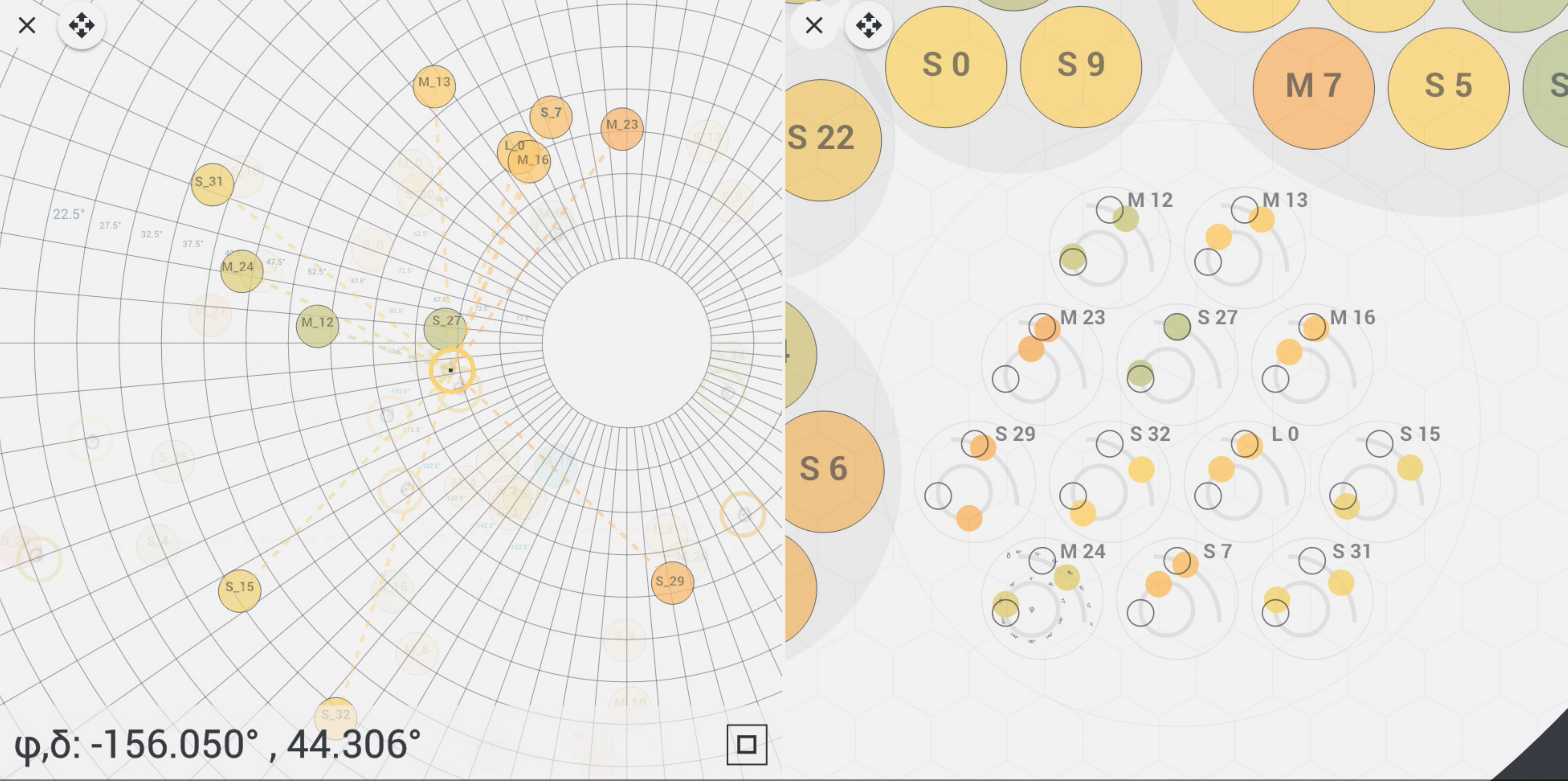}}
  \end{minipage}\hfill \\
  \begin{minipage}[c]{1\textwidth}
    \centering\subfloat[]{\label{FIG_arrPos_2}\includegraphics[trim=0mm -30mm 0mm -25mm,clip,width=.68\textwidth]{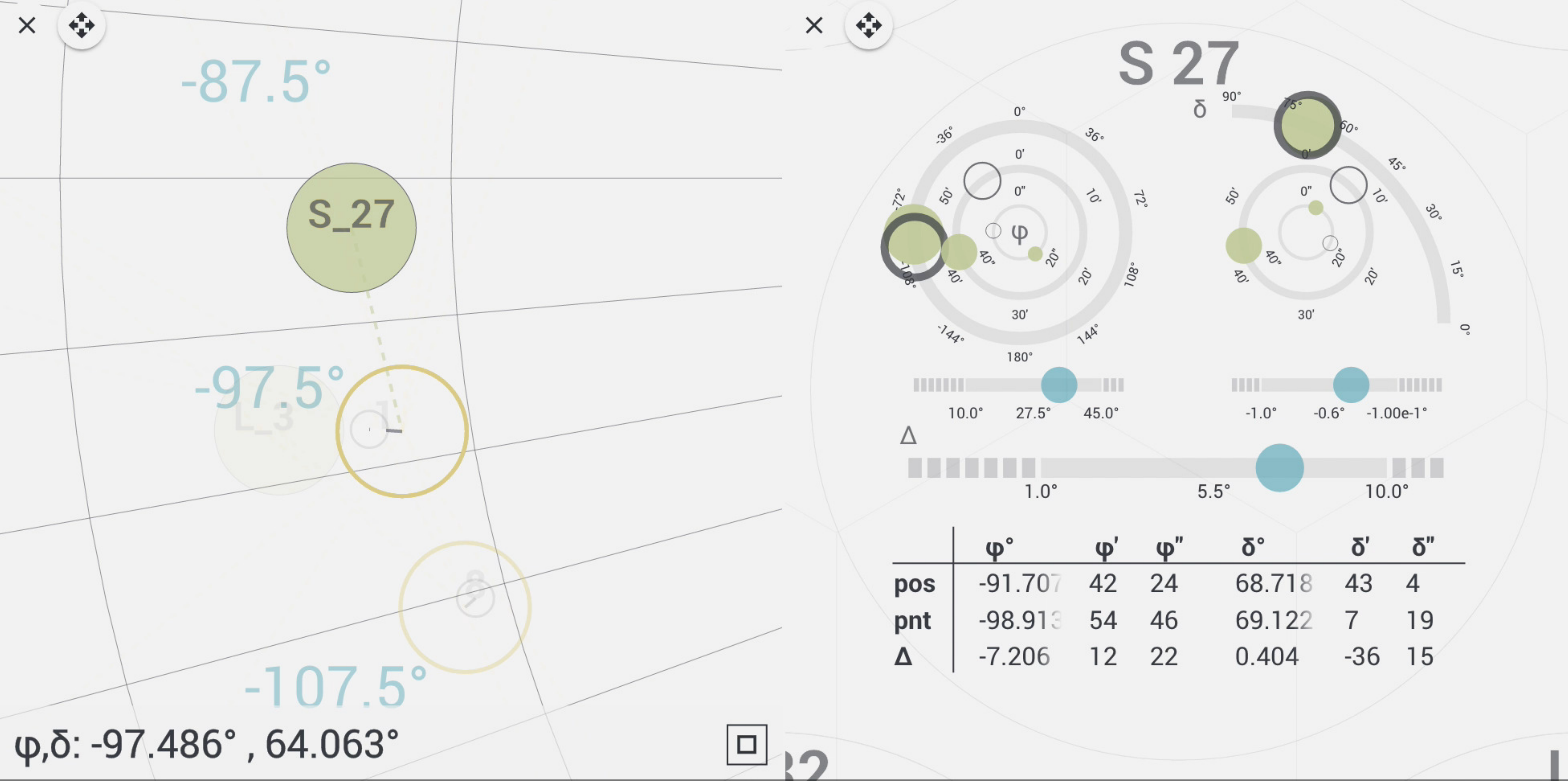}}
  \end{minipage}\hfill \\
  \vspace{15pt}
  \caption{\label{FIG_arrPos}
      Synchronized panels, showing the positions on the sky at which telescopes point (left), and
      the grouping of telescopes into sub-arrays (right), as described in the text.
      The zoom factors for the panel on the right are~$1$ in \Subref{FIG_arrPos_0}, $2$ in~\Subref{FIG_arrPos_1},
      and~$\sim14$ in~\Subref{FIG_arrPos_2}.
  }
\end{center}
\end{figure*}

\Autoref{FIG_arrPos} presents another telescope-centric view.
In this case, two synchronized panels are provided side by side.
The panel on the left shows the positions on the sky, at which telescopes point. The polar plot has two coordinates.
The angular coordinate, denoted by $\varphi$, represents the azimuth on the sky; it spans the
range from~$0^{\circ}$ (at the top-most position of the circle)
to~$180^{\circ}$ (for the right-hand side), and~$0^{\circ}$ to~$-180^{\circ}$ (for the left-hand side).
The radial coordinate, denoted by $\delta$, indicates the elevation, with values between~$0^{\circ}$
(at the edge of the figure) and~$90^{\circ}$ (on the centre).
Telescopes are represented by full circles. Each one has an intended target position,
indicated by rings, where faint dashed lines connect telescopes and targets.
The display is updated in real-time, showing the movement of telescopes from their initial
pointing, towards their respective target positions.

The panel on the right shows a logical grouping of telescopes, as opposed to the
pseudo-geographic display of \autoref{FIG_arrZoom}. In this case, telescope positions
indicate association to a sub-array, which here
stands for a group of telescopes which all point at the same target on the sky.
This type of visualization, called an \textit{enclosure diagram}, is a hierarchical
nested layout of elements. In this case, grey-shaded circles
represent sub-arrays, with coloured circles standing for telescopes.

Both the panel on the left and that on the right incorporate semantic zooming, as
indicated by comparing \autorefs{FIG_arrPos_0}, \ref{FIG_arrPos_1}, and~\ref{FIG_arrPos_2}.
For the left panel, the behaviour is a simple scaling of elements.
As one zooms in or out, the size of telescopes and targets changes
very slowly with respect to the changing scale of the coordinate system.
For the panel on the right, the semantic zoom includes a threshold transition
at zoom factor~$\sim2$, as shown in \autoref{FIG_arrPos_1}. In this example,
each telescope is represented by a circle and an arc; these respectively
indicate the $\varphi$ and $\delta$ coordinates of the telescope and its target,
using the same visual language as for the panel on the left.
Another threshold transition occurs
at zoom factor~$\sim14$, as presented in \autoref{FIG_arrPos_2}. 
In this instance, the purpose of the visualization is to provide
detailed information, regarding the coordinates at which the telescope is pointing.

The two panels are synchronized using brushing and linking,
following the principles of coordinated multiple views~\cite{North00}.
In \autoref{FIG_arrPos_1}, a sub-array
is selected on the right panel; this causes the associated
elements in the panel on the left to become highlighted.
Such behaviour allows to quickly identify in the display on the left,
which telescopes are assigned to which sub-array.
In \autoref{FIG_arrPos_2}, the single selected telescope is
similarly the only element in focus, both on the right panel and on the left.
The interplay between the two panels illustrated here also serves
to demonstrate an important feature of panel-synchronization. Namely,
that synchronization need not necessarily work both ways. As in
this case, user interactions with the panel on the right affect the panel on
the left by focusing specific elements, but the reverse does not hold.

\begin{figure*}[htbp]
\begin{center}
\vspace{-5mm}
  \begin{minipage}[c]{1\textwidth}
    \centering\subfloat[]{\label{FIG_weather_0}\includegraphics[trim=0mm -35mm 0mm -25mm,clip,width=.96\textwidth]{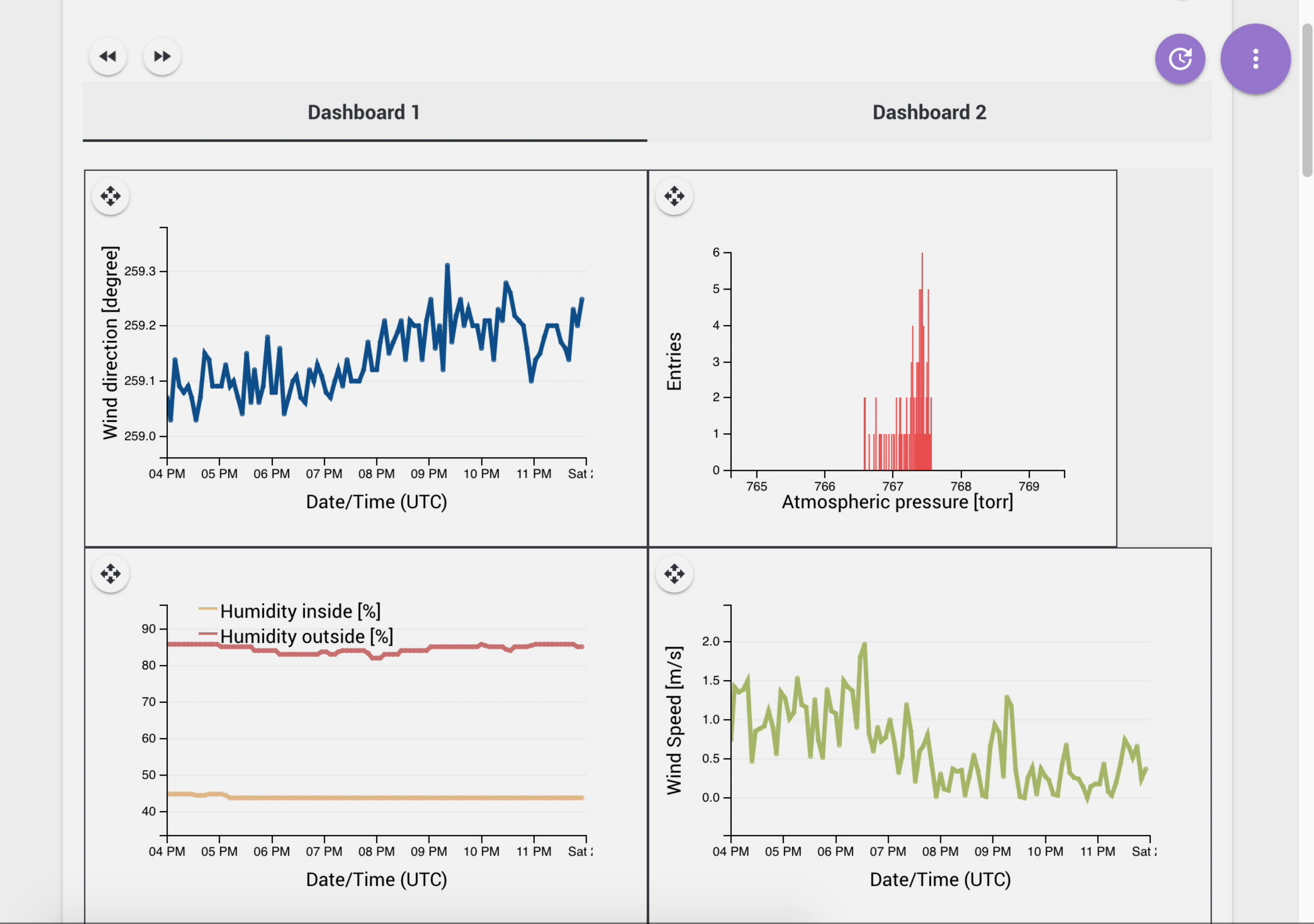}}
  \end{minipage}\hfill \\
  \begin{minipage}[c]{0.5\textwidth}
    \centering\subfloat[]{\label{FIG_weather_1}\includegraphics[trim=0mm -80mm 0mm -25mm,clip,width=.92\textwidth]{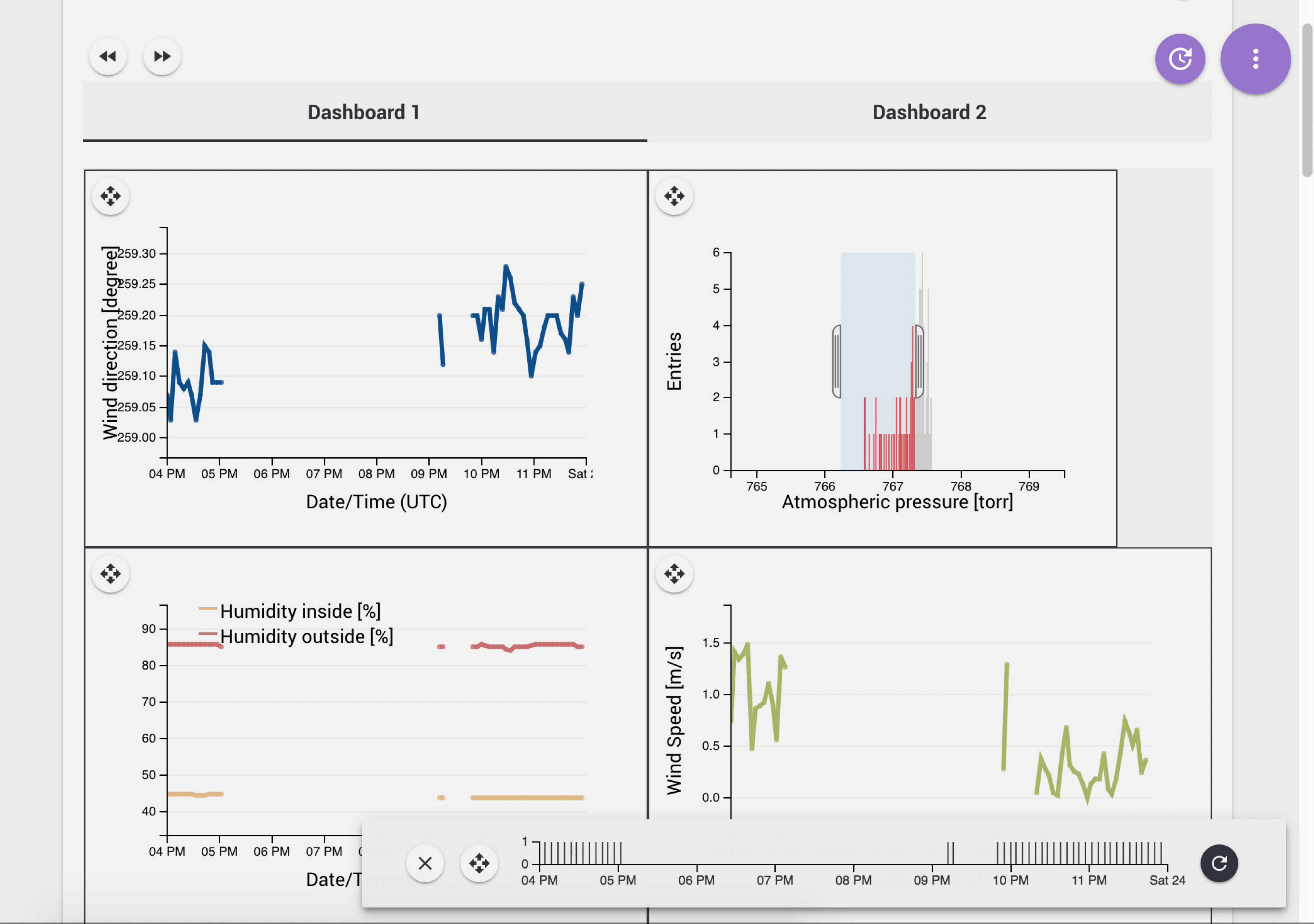}}
  \end{minipage}\hfill
  \begin{minipage}[c]{0.5\textwidth}
    \centering\subfloat[]{\label{FIG_weather_2}\includegraphics[trim=0mm -80mm 0mm -25mm,clip,width=.92\textwidth]{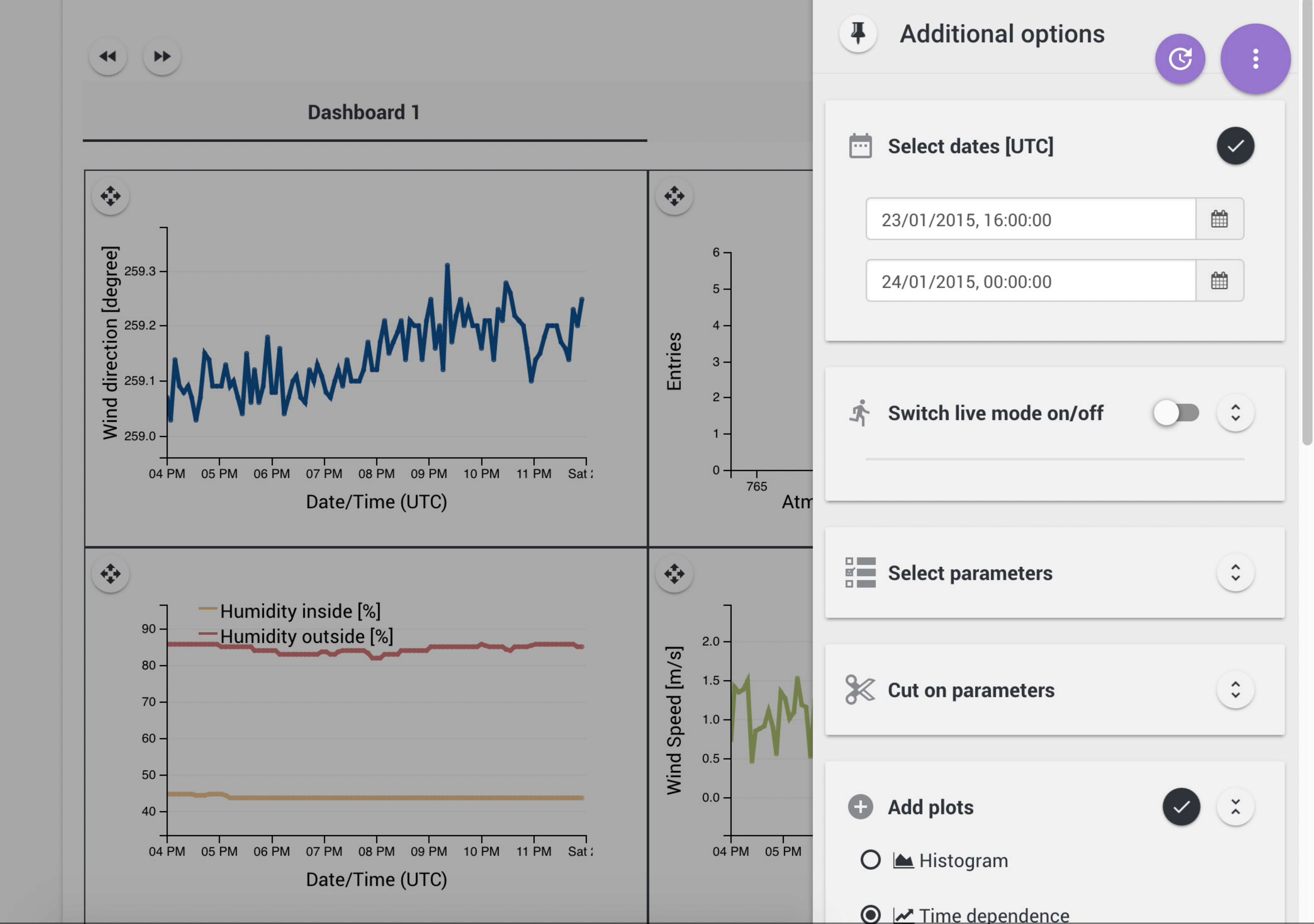}}
  \end{minipage}\hfill
  %

  %
  \vspace{15pt}
  \caption{\label{FIG_weather}
    Example of a monitoring panel, featuring data extracted from the database of a weather monitoring station.
    These include measurements of the
    humidity inside and outside of the weather station, of the speed and direction of the wind near
    the station, and of the atmospheric pressure.
    \Subref{FIG_weather_0}~:~A selection of data monitoring plots.
    \Subref{FIG_weather_1}~:~Illustration of the synchronization of plots. A selected range of values
    of the atmospheric pressure parameter (shaded in blue), affects the other plots, by excluding data outside
    of the chosen scope.
    \Subref{FIG_weather_2}~:~A side-menu with various control options may be overlaid on top of the plots,
    upon clicking the purple button on the top right corner.
    The option also exists (not shown), of pinning the menu alongside the plots instead.
  }
\end{center}
\end{figure*}

\Autoref{FIG_weather} shows three views of a panel featuring monitoring plots.
In this example, the data are extracted from the database of a
weather monitoring station. They include measurements of the
humidity inside and outside of the weather station, of the speed and direction of the wind near
the station, and of the atmospheric pressure.
The monitoring plots are placed inside a container, designated as a \textit{Dashboard}. Plots
may be resized and dragged within the Dashboard, or placed in separate Dashboards.
The plots are generated using \dcs.
They are correlated (a feature of \crsfilt), as shown in, \eg \autoref{FIG_weather_1}.
In this case, the user has selected a range of values of the atmospheric pressure parameter (marked by the blue brush).
The other plots reflect the selection, by showing only those measurements
which have a corresponding atmospheric pressure within the chosen range.
A side-menu may be overlaid on top of the plots, as shown in \autoref{FIG_weather_2}.
Several control options exist. To name a few, the user may add new plots;
the user may select new dates for querying the database;
and the user may switch to a \textit{live-mode},
where real-time data are updated every few seconds.

The main purpose for implementing this visualization was to
create a practical example for a monitoring panel. In addition,
it was also used as a platform to conduct tests of our prototype.
These included the following:
testing the performance of the \dcs library for generating synchronized plots;
testing the performance of an interface between the \python web-server and an external database;
splitting data processing tasks between the \python server and the web browser client.
The results of these tests have so far been encouraging; they have indicated that a web-based
framework is suitable for the next stages of development of the operator GUI.

\section{SUMMARY}\label{sec_summary}
%
The planned Cherenkov Telescope Array will be comparably more complex than
existing IACT experiments.
This poses new challenges for creating an effective graphical user interface
for the operator of the array.

The development of the GUI for CTA follows the successful
working model of the ALMA experiment. The design process brings together experienced telescope
operators and astroparticle physicists on the one hand, and experts from the field of
Human-Computer Interaction on the other. The relevant stakeholders have
so far conducted two participatory design workshops, in addition to other
face-to-face meetings. The outcome of the work to date is:
a refined definition of the scope of the GUI and of the requirements on its performance;
a preliminary list of GUI panels;
identification of a set of relevant data visualization techniques.

A preliminary prototype of several GUI panels has been implemented.  It is
based on Web technologies, incorporating a \python web server,
\webSock and graphics generated with the \djs \jvs library.
The prototype illustrates semantic zooming and coordinated multiple views, and serves
for performance testing of the proposed technology.

\acknowledgments 
 
We would like to thank Caroline Appert, Antonio Cabrera, Francesco Dazzi,
Valentina Fioretti, Stafano Gabici, Markus Garczarczyk, Rosa Macias and the members of the
ACTL team, for their helpful comments and insights.

\bibliography{bib} 
\bibliographystyle{spiebib} 

\end{document}